%% file: document_revision.tex
\documentclass{bioinfo}
\copyrightyear{2005}
\pubyear{2005}

\usepackage{amssymb,amsfonts,amsmath}
\usepackage{graphicx}
\usepackage{caption}
\usepackage{subfig}
\usepackage{multirow}
\usepackage{fullpage}
\usepackage{booktabs}

\begin{document}
\graphicspath{{figures_revision/}}
\firstpage{1}

\title[Multi-Marker Mixed Model for Association Mapping]{LMM-Lasso: A Lasso Multi-Marker Mixed Model for Association Mapping with Population Structure Correction}
\author[B. Rakitsch \textit{et~al}]{Barbara Rakitsch$^{1,2,\ast}$,
Christoph Lippert$^{1,2,\ast}$,
Oliver Stegle$^{1,2,\ast}$,
Karsten Borgwardt$^{1,2,3}$}
\address{
\bf{1} Max Planck Institute for Developmental Biology, T\"ubingen, Germany
\\
\bf{2} Max Planck Institute for Intelligent Systems, T\"ubingen, Germany
\\
\bf{3} Eberhard Karls Universit\"at T\"ubingen, Germany
\\
}

\input{utils}

\history{Received on XXXXX; revised on XXXXX; accepted on XXXXX}

\editor{Associate Editor: XXXXXXX}

\maketitle

\begin{abstract}

\section{Motivation:}
Exploring the genetic basis of heritable traits remains one of
the central challenges in biomedical research.
In traits with simple mendelian architectures,
single polymorphic loci explain a significant fraction of the phenotypic variability.
However, many traits of interest appear to be subject to
multifactorial control by groups of genetic loci.
Accurate detection of such multivariate associations is non-trivial
and often compromised by limited power.
At the same time, confounding influences such as population structure
cause spurious association signals that result in false positive
findings if they are not accounted for in the model.

\section{Results:}
We propose LMM-Lasso, a mixed model that allows for both
multi-locus mapping and correction for confounding effects.
Our approach is simple and free of tuning parameters, effectively
controls for population structure and scales to genome-wide datasets.
LMM-Lasso simultaneously discovers likely causal variants
and allows for multi-marker based phenotype prediction from genotype.
We demonstrate the practical use of LMM-Lasso in genome-wide association studies in \emph{Arabidopsis thaliana} and linkage mapping in mouse, where
our method achieves significantly more accurate phenotype prediction
for $91\%$ of the considered phenotypes.
At the same time, our model dissects the phenotypic variability into
components that result from individual SNP effects and population
structure.
Enrichment of known candidate genes suggests that the individual associations
retrieved by LMM-Lasso are likely to be genuine.

\section{Availability:}
Code available under XXX.

\section{Contact:} \href{ \{rakitsch, clippert, stegle\}@tuebingen.mpg.de}{ \{rakitsch, clippert, stegle\}@tuebingen.mpg.de}
\end{abstract}

\section{Introduction}
While many quantitative traits in humans, plants and animals have been
observed to be heritable, a comprehensive understanding of the
underlying genetic architecture is still missing.
In some cases genome-wide association studies and linkage
mapping have already revealed individual causal variants that
control trait variability; for example, genetic mapping yielded
insights into the genetic architecture of global-level traits in
plants~\cite{Atwell2010} and
mouse~\cite{valdar2006genome}, as well as the risks for important human
diseases such as type 2 diabetes~\cite{craddock2010genome}.
Nevertheless, the statistical analysis of these genetic data has
proven to be challenging, not
least because single genetic variants rarely explain larger fractions
of phenotype variability, and hence, individual effect sizes are
small~\cite{mccarthy2008genome,
  mackay_genetics_2009}.
An inherent limitation of power to map weak effects is due to
confounding relatedness between samples.
Population structure can induce false association patterns with large
numbers of loci being correlated with the phenotype. 
To understand the true genetic architecture of complex traits, it is
necessary to address both of these challenges, taking population
structure into account and joint modeling of true multifactorial
associations.   

If multiple variants contribute to phenotype variation in an
additive fashion, simple methods that assess the significance
of individual loci independently are likely to fall short:
masking effects between causal SNPs can limit mapping power, with
relevant loci not reaching genome-wide significance
levels~\cite{mccarthy2008genome}.
These shortcomings have been widely addressed in multivariate
regression, explicitly modeling the additive effect of multiple SNPs.

The corresponding methods either fit sparse predictors of all genome-wide SNPs,
using a shrinkage prior or employ
stepwise forward selection~\cite{yang2012conditional}.
Applying a Laplace prior leads to the Lasso~\cite{li2011bayesian},
and related priors have also been considered~\cite{balding-2008-1}.
With the same ultimate goal to capture the genetic effects of groups
of SNPs, variance component models have recently been proposed to
quantify the heritable component of phenotype variation explainable by
an excess of weak effects~\cite{visscher-2010-1}.

Population structure induces spurious correlations between genotype
and phenotype, complicating the genetic analysis.
A major source of these effects can be understood as deviation from
the idealized assumption that the samples in the study population are
unrelated.
Instead, population structure in the sample is difficult to avoid and
even in a seemingly stratified sample, the extent of hidden structure
cannot be ignored~\cite{newman2001importance}.
Models that account for the presence of such structure are routinely
applied and have been shown to greatly reduce the impact of this
confounding source of variability.
For instance, EIGENSTRAT builds on the idea of extracting the major axes
of population differentiation using a PCA decomposition of the
genotype data~\cite{price2006principal}, and subsequently including them into the model
as additional covariates.
Linear mixed models~\cite{yu2006unified,kang2008efficient,zhang2010,Kang2010,fastlmm}
provide for more fine-grained control by modeling the contribution of
population structure as a random effect, providing for an effective 
correction of family structure and cryptic relatedness. 

\EDIT{
While both, correction for population structure and joint mapping of
multiple weak effects, have been addressed in isolation, few existing
approaches are capable of addressing both aspects jointly. 
In line with EIGENSTRAT, the authors
of~\cite{balding-2008-1,li2011bayesian} add principal components to  
the model to correct for population structure. 
In parallel to our work, Segura et. al~\cite{bjarni2012} have proposed
a related multi-locus mixed model approach, however employing
step-wise forward selection instead of using the Lasso.
}

 Here, we propose a novel analysis approach that combines multivariate
association analysis with accurate correction for population
structure.
Our method allows for joint identification of sets of loci that
individually have small effects and at the same time accounts for
possible structure between samples.
This joint modeling explains larger fractions of the total phenotype
variability while dissecting it in variance components specific to
individual SNP effects and population effects.

Our approach bridges the advantages of linear mixed models with
Lasso regression, hence, modeling complex genetic effects while
controlling for relatedness in a comprehensive fashion.
The proposed LMM-Lasso is conceptually simple, computationally
efficient and scales to genome-wide settings.
Experiments on semi-empirical data show that the rigorous combination
of Lasso and mixed modeling approaches yields greater power to detect
true causal effects in a large range of settings.
In retrospective analyses of studies from \emph{Arabidopsis} and
mouse, we show that through joint modeling of population structure and
individual SNP effects, LMM-Lasso results in superior models of the genotype to
phenotype map.
These yield better quantitative predictions of phenotypes while
selecting only a moderate number of SNP with individual effects.
Additional evidence of the effects uncovered by LMM-Lasso likely being
real is given by an enrichment analysis, suggesting that the hits
obtained are often in the vicinity of genes with known 
implication for the phenotype.

\section{Multivariate linear mixed models}
Our approach builds on multivariate statistics, explaining the
phenotype variability by a sum of individual genetic effects and
random confounding variables. 
In brief, the phenotype of $m$ samples $\bfy = (y_1, \ldots, y_m)$ is
expressed as the sum of $n$ SNPs $\bfS =(\mathbf s_1, \ldots, \mathbf
s_n)$
\begin{align}
\label{eq:model}
  \bfy =\underbrace{\sum_{j=1}^{n} \beta_j \bfs_j}_{\text{genetic
      factors}} +\underbrace{\bfu}_{\text{confounding}} + \underbrace{\bpsi}_{\text{noise}}.
\end{align}
Here, $\bpsi$ denotes observation noise and $\bfu$ are confounding
influences.
Confounding influences in genetic mapping are typically not directly observed,
however their Gaussian covariance $\bfK$ can in many cases be
estimated from the observed data. 
To account for confounding by population structure, $\bfK$ can be reliably estimated
from genetic markers, for example using the realized relationship matrix which captures the overall
genetic similarity between all pairs of samples~\cite{hayes-2009-1}.
Similarly, in genetic analyses of gene expression, $\bfK$ can be fit
to capture and correct for the confounding effect of gene expression
heterogeneity~\cite{listgarten2010correction,fusi2012joint}.
Marginalizing over the random effect $\bfu$ results in a Gaussian
marginal likelihood model~\cite{kang2008efficient} whose covariance
matrix accounts for confounding variation and observation noise.  

The resulting mixed model is typically considered in the context of
single candidate SNPs, i.e. restricting the sum in
Eq.~\eqref{eq:model} to a particular SNP while ignoring all
others~\cite{yu2006unified,kang2008efficient,zhang2010,Kang2010,fastlmm}.
While computationally efficient and easy to interpret, this
independent analysis can be compromised by complex genetic
architectures with some genetic factors masking
others~\cite{platt2010conditions}. 
\EDIT{
Some improvement can be achieved by step-wise regression or forward
selection, which has recently been extended to the mixed model
framework~\cite{yang2012conditional, bjarni2012}.
However as any step-wise procedure in general, these approaches are
prone to retrieving local optima as the order in which SNP markers are
added matters.
}
As an alternative,  we propose an efficient approach to carry out
joint inference in the model implied by Eq.~\eqref{eq:model}.
Our approach assesses all SNPs at the same time while accounting for
their interdependencies and without making any assumptions on their
ordering.
To allow for applications to genome-wide SNP data, we place a
Laplacian shrinkage prior over the fixed effects~$\beta_i$, assigning
zero effect size to the majority of SNPs as done in the
classical Lasso~\cite{tibshirani-1996-lasso}. 

We call this approach LMM-Lasso as it combines the
advantages of established linear mixed models (LMM) with sparse Lasso
regression. 
The resulting model allows for dissecting the explained phenotype
variance into a component due to individual SNP effects and effects
caused by confounding structure. 

\subsection{Linear mixed model Lasso}
Let $\bfS$ denote the $m \times n$ matrix of $n$ SNPs for $m$
individuals, $\bfs_j$ is then the $m \times 1$ vector representing SNP 
$j$. 
We model the phenotype for $m$ individuals, $\bfy=(y_1,\dots,y_m)$ as
the sum of genetic effects $\beta_j$ of SNPs $\bfs_j$ and confounding
influences $\bfu$ (see Eq.~(\ref{eq:model})). 
The genetic effects are treated as fixed effects, whereas the
confounding influences are modeled as random effects.
The genetic effect terms are summed over genome-wide polymorphisms,
where the great majority of SNPs has zero effect size,
i.e. $\beta_j=0$, which is achieved by a Laplace shrinkage prior on
all weights. 
The random variable $\bfu$ is not observed directly.
Instead, we assume that the distribution of $\bfu$ is Gaussian with
covariance $\bfK$, $\bfu \sim \mathcal{N}(0,\sigma_g^2 \bfK)$.

\EDIT{
Assuming Gaussian noise, $\bpsi \sim \mathcal{N}(0,\sigma_e^2
  \unit)$, and marginalizing over the random variable $\bfu$, we can
  write down the conditional posterior distribution over the weight
  vector $\bbeta$: 
  }
\begin{align}
\label{eq:posterior_beta}
\nonumber
  &p(\bbeta \given \bfy,  \bfS, \bfK, \sigma_g^2,\sigma_e^2, \lambda) 
  \propto\\
  & \;\;\;\;\;\;\;\;\;
  \underbrace{\normal{\bfy}{\sum_{j=1}^{n} \beta_j \bfs_j, \sigma_g^2 \bfK +
    \sigma_e^2 \unit}}_{\text{marginal likelihood}}
     \underbrace{\prod_{j=1}^{n} e^{-\frac{\lambda}{2} |\beta_j|_{1}}}_{\text{prior}}.
\end{align}
Here, $\lambda$ denotes the sparsity hyperparameter of the Laplace
prior, $\sigma_e^2$ is the residual noise variance and $\sigma_g^2$
denotes the variance of the random effect components.

\subsection{Parameter inference}
\label{sec:corr-popul-struct}

Learning the hyperparameters $\bTheta=\{ \lambda, \sigma_g^2 ,
\sigma_e^2 \}$ and the weights $\bbeta$ jointly is a hard non-convex
optimization problem.
Here, we propose a combination of fitting some of these parameters on
the null model with the individual SNP effects excluded and reduction to a
standard Lasso regression problem. 

\paragraph{Null-modell fitting}
To obtain a practical and scalable algorithm, we first optimize
$\sigma_g^2,\sigma_e^2$ by Maximum Likelihood under the null
model, ignoring the effect of individual SNPs.
The analogous procedure is widely used in single-SNP mixed models, 
and has been shown to yield near-identical result to an exact
approach~\citep{Kang2010}. 
To speed up the computations needed, we optimize the ratio of the
random effect and the noise variance, $\delta=
{\sigma_e^2}/{\sigma_g^2}$, which can be optimized efficiently by
using computational tricks proposed elsewhere~\cite{fastlmm}:
\begin{align}
  p(\bbeta \given \bfy, \bfS, \bfK, \sigma_g^2,\delta, \lambda) \propto
  \normal{\bfy}{\sum_{j=1}^{n} \beta_j \bfs_j, \sigma_g^2 (\bfK +
    \delta \unit)} \prod_{j=1}^{n} e^{-\frac{\lambda}{2} |\beta_j|_{1}}.
\end{align}
Briefly, we compute the eigendecomposition of the covariance $\bf K=
\bfU   \diag(\bfd) \bfU^{\T}$ which can be used to rotate the data
such that the covariance matrix of the normal distribution is isotropic.
We carry out one-dimensional numerical optimization of the marginal
likelihood (Eq.~\eqref{eq:posterior_beta}) with respect to $\delta$,
whereas $\sigma_g^2$ can be optimized in closed form in every
evaluation.   
 
\paragraph{Reduction to standard Lasso problem}
Having fixed $\delta$, we use the eigendecomposition of $\bfK$ again
to rotate our data such that the covariance matrix becomes isotropic: 

\begin{align}
\label{eq:model2}
  p(\bbeta \given \tilde{\bfy} , \tilde{\bfS}, \bfK, \sigma_g^2, \lambda) \propto
  \normal{\tilde{\bf y}}{\sum_{j=1}^{n} \beta_j  \tilde{ \bfs}_j, \sigma_g^2  \unit} \prod_{j=1}^{n} e^{-\frac{\lambda}{2} |\beta_j|_{1}}
\end{align}
Here, $\tilde {\bfS}$ denote the rotated and rescaled genotypes and $\tilde {\bfy}$ the respectively phenotypes:
\begin{eqnarray*}
\bf{\tilde S} = (\diag(\bfd)+\delta\bfI)^{-\frac12}\bf {U^T S}, &&
\bf{\tilde y} =  (\diag(\bfd)+\delta\bfI)^{-\frac12}\bf {U^T y}.
\end{eqnarray*}
Using this transformation, the task of determining the most probable
weights in Eq.~\eqref{eq:model2} is now equivalent to the Lasso
regression model, since maximizing the posterior with respect to
$\bbeta$ is equivalent to minimizing the negative log of
Eq. \eqref{eq:model2}: 
\begin{align}
\min_{\bbeta}  \frac1{\sigma_g^2} || {\tilde \bfy} - \tilde{\bfS}
\bbeta ||_{2} +  \lambda \| \bbeta \|_1 \nonumber. 
\end{align}
A related algorithm for combining random effects with the Lasso has
been proposed in~\cite{schelldorfer-2011-1}, which includes a
generalized linear mixed models with $\ell_1$-penalty at the cost of 
higher computational complexity.
An appropriate setting of $\lambda$ can be found by cross-validation
to maximize the overall predictive performance or stability selection~\cite{buehlmann-ss-2010-1}.

\EDIT{
The computational efficiency of the two-stage procedure proposed here
depends on the approximation to fit $\delta$ on the null model,
allowing for the reduction of the problem to standard Lasso
regression. 
For univariate single-SNP mixed models, efficient optimization of
$\delta$ for each SNP can be done by recently proposed
computational tricks~\cite{fastlmm,Stephens2012}.    
Unfortunately, these techniques cannot be directly applied in the
multivariate setting.  
In principle it is possible to extend the cross-validation to optimize
over pairs ($\delta, \lambda$).
However, this remains impracticable for most datasets due to the
additional computational cost implied and hence we consider optimizing
$\delta$ on the null model in the experiments~\cite{Kang2010}.    

\subsection{Phenotype prediction}
\label{sec:pred-this-model}
Given a trained LMM-Lasso model on a set of genotype and phenotypes, 
we can predict the unobserved phenotype of test individuals.
The predictive distribution can be derived by conditioning the joint
distribution over all individuals on the  training
individuals~\cite{rasmussen_gaussian_2006}, resulting in a Gaussian
predictive distribution $p( \bfy^{\star} 
\given \bfy, {\bf S^{\star}}, 
\bfS) = \normal{\bfy^{\star}}{\bmu^{\star},\bSigma^{\star}}$, with 
\begin{align}
  \label{eq:predictive_distribution}
  \nonumber
  \bmu^{\star} &=& \underbrace{{\bf S^{\star} }\bbeta}_{\text{Lasso prediction}} + \underbrace{\bfK_{\bfS^{\star}\bfS} (\bfK + \delta
 \bfI)^{-1} (\bfy-\bfS\bbeta)}_{\text{Random effect prediction}} \\
  \bSigma^{\star} &=&  \sigma_g^2 (\bfK_{\bfS^{\star}\bfS^{\star}} + \delta\bfI)  -
  \sigma_g^2 \bfK_{\bfS^{\star}\bfS}  (\bfK + \delta \bfI)^{-1} \bfK_{\bfS\bfS^{\star}}.
\end{align}
The mean prediction is a sum of contributions from the Lasso component
and the random effect part, which is similar to
BLUP~\citep{robinson1991blup}. 
The matrix $\bfK_{\bfS^{\star}\bfS} $ denotes the covariance matrix between
the test individuals $\bf S^{\star}$ and the train individuals $\bf S$,
$\bfK_{\bfS^{\star}\bfS^{\star}} $ is the covariance matrix between all test
individuals and $\bfK := \bfK_{\bfS\bfS} $ is the covariance matrix
between all training individuals, which with slight abuse of notation are
denoted by their genetics $\bfS$.
}

\subsection{Choice of the random effect covariance to account for
  population structure}
\label{sec:choice-random-effect}
\EDIT{
Depending on the application, the random effect covariance $\bfK$ can
be chosen in a variety of ways.
Here, we discuss specific options to account for population structure.

\paragraph{Choice of genetic similarly matrix}
For the identity by descent matrix (IBD), an entry is defined as the
predicted proportion of the genome that is identical by descent given
the pedigree information.  
In contrast, the identity by state matrix (IBS) simply counts the
number of loci on which the samples agree, where as the realized
relationship matrix (RRM) is calculated as the linear kernel between
the SNPs~\cite{hayes-2009-1}. 
In subsequent experiments, we have used the realized relationship
matrix. An example for the RRM-matrix derived from the \textit{Arabidopsis 
thaliana} dataset is given in Figure S1.
}

\paragraph{Realized relationship matrix and relationship to Bayesian 
  linear regression}
\EDIT{From a Bayesian perspective, employing the realized relationship
  matrix as the covariance matrix is equivalent to integrating over
  all SNPs in a linear additive model with an independent Gaussian
  prior over the weights 
  $\normal{\bbeta}{{\bf{0}},\sigma_g^2 \unit}$
  ~\cite{goddard_2009-1}.
The choice of a Gaussian prior leads to a dense posterior
distribution, reflecting the \emph{a priori} belief that a large fraction of
SNPs jointly contribute to phenotype variability. 
This prior choice is in sharp contrast to the generally accepted
opinion that most SNPs are not causal. 

Thus, choosing this particular covariance matrix $\bf K$ can be
regarded as modeling genetic effects that are confounded due to population
structure or are small additive infinitesimal effects, whereas single
SNPs that have a sufficiently large effect size are directly included
in the Lasso of model.
}

\subsection{Scalability and runtime}
\label{sec:scalability-runtime}
\EDIT{
The appeal of the LMM-Lasso is a runtime performance comparable 
the standard LASSO. 
The difference is a one-time off cubic cost for the decomposition of
the random effect matrix $\bfK$ to rotate the genotype and phenotype
data (see Section~\ref{sec:corr-popul-struct}).

To demonstrate the applicability to genome-wide datasets, we have
empirically measured the runtime for computing the complete 
path of sparsity regularizers on the synthetic dataset, consisting of
1,196 plants and 213,624 SNPs.
On a single core of a Mac Pro (3GHz, 12 MB L2-Cache, 16GB Memory), the
Lasso required 145 minutes CPU time and the LMM-Lasso 146 minutes of
CPU time. 

If needed, the runtime of LMM-Lasso could be improved in several ways.
First, if the number of samples is large (m $> 10^5$), the runtime is
dominated by the decomposition of $\bfK$ and rotating the data for
the optimization of $\delta$.  
As shown in~\cite{fastlmm}, reducing the covariance $\bfK$ to a
low-rank representation calculated from a small subset of $n_s$ SNPs,
yields very similar results while reducing the 
runtime from $O(m^2n)$ to $O(mn_{s}^2)$. 
Second, the runtime of the $\ell_1$-solver is heavily dependent on the
optimization method used. 
Fortunately, the development of new and efficient $\ell_1$-solvers is
still an active area of research. 
New approaches include parallelized coordinate descent
algorithms~\cite{bradley2011} and screening tests that are able to
prune away SNPs that are guaranteed to have zero
weights~\cite{xiang-2011-1}, avoiding to load the complete genotype
matrix into the working memory. 
}
\begin{methods}

\section{Methods and Material}
\label{sec:datasets}
\subsection{{\it Arabidopsis thaliana}}
\label{sec:it-arabidopsis}
We obtained genotype and phenotype data for up to 199 accessions of
{\it Arabidopsis thaliana} from~\cite{Atwell2010}. Each genotype
comprises 216,130 single nucleotide polymorphisms per accession. 
We study the group of phenotypes related to the flowering time of the
plants. We excluded phenotypes that were measured for less than 150
accessions to avoid possible small sample size effects, resulting in a
total of 20 flowering phenotypes that were considered.
The relatedness between individuals ranges in a wide spectrum leading
to a complex population structure~\cite{platt-2010-1}. 

\subsection{Mouse inbred population}
We also obtained genotype and phenotype data for 1,940 mice from a
multi-parent inbred population~\cite{valdar2006genome}. 
Each individual genotype comprises of 12,226 single nucleotide
polymorphisms.  
All mice were derived from eight inbred strains and were crossed to
produce a heterogenous stock. 
The phenotypes span a large variety of different measurements ranging
from biochemical to behavioral traits.
Here, we focused on 273 phenotypes which have numeric or binary
values. 

\subsection{Semi-empirical data}
\label{sec:semi-empirical-data}
To assess the accuracy of alternative methods for variable selection, 
we considered a semi-empirical example based on the extended \emph{A. thaliana}
dataset~\cite{bergelson-2012-1} consisting of 1196 plants.
We considered real phenotype data to obtain realistic background
signal that is subject to population structure. 
In addition to this empirical background, we added simulated
associations with different effect sizes and a range of complexities
of the genetic models.
For full details of the simulation procedure and the evaluation of 
associations recovered by different methods, see supplementary text.

\subsection{Preprocessing}
\label{sec:preprocessing}
We standardized the SNP data which has the effect that SNPs with a
smaller MAF have a larger effect size as reported in
\cite{amos_shifting_2008-1}. 
On the phenotypes, we performed a Box-Cox
transformation~\cite{saskia_boxcox-1992-1} and subsequently
standardized the data. 

\subsection{Model Selection}
\label{sec:model-selection}
\EDIT{
Variation of the model complexity of Lasso Methods can either be done
by choosing the number of active SNPs or equivalently by varying the
hyperparameter $\lambda$ explicitly. 
For the benefit of direct interpretability, we chose to vary the
number of active SNPs.  
For a fixed number of selected SNPs, we find the corresponding
hyperparameter $\lambda$ by a combination of bracketing and bisection
as done in~\cite{wu_gwaslars-2009-1}. 

To select which of these Lasso-model is most suitable, we consider
alternative strategies, depending on the objective.
\begin{enumerate}
\item \textbf{Phenotype prediction}
  To predict phenotypes, we use 10-fold cross-validation.
  We split the data randomly into 10 folds. 
  Each fold is once picked as test dataset, with all other folds being
  used for training the model.
  The model is selected to maximize the explained variance on the test set.
  In this comparison, we considered models with different numbers of
  SNPs, varying from
  $\{0,1,2,\ldots,10,20,30,\ldots,100,150,200,250\}$ with the
  additional constraint that the number of active SNPs shall not exceed 
  the number of samples.

\item \textbf{Variable selection}
  To assess the significance of individual features, we consider
  stability Selection~\cite{buehlmann-ss-2010-1}. 
  Here, we fix the number of active SNPs to $20$ and draw randomly
  $90\%$ of the data $100$ times. 
  Significance estimates can be deduced from the selection frequency
  of individual SNPs (see~\cite{buehlmann-ss-2010-1}).
  
  To obtain a complete ranking of features, as used to evaluate models
  in the simulation study, we use the LASSO regularization path and
  rank features by the order of inclusion into the model.
\end{enumerate}
}

\section{Results}
\subsection{Semi-empirical setting with known ground truth}

\begin{figure}[!ht]
  \begin{center}
    \includegraphics[width=0.45\textwidth]{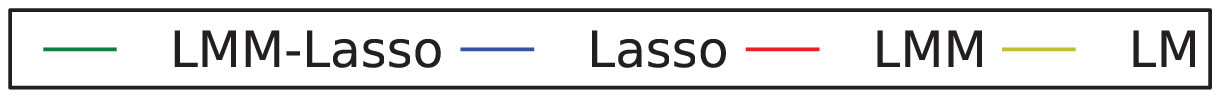} \\
        \vspace{-0.5cm}
         \subfloat[][{Precision/Recall}]{
    \includegraphics[width=0.225\textwidth]{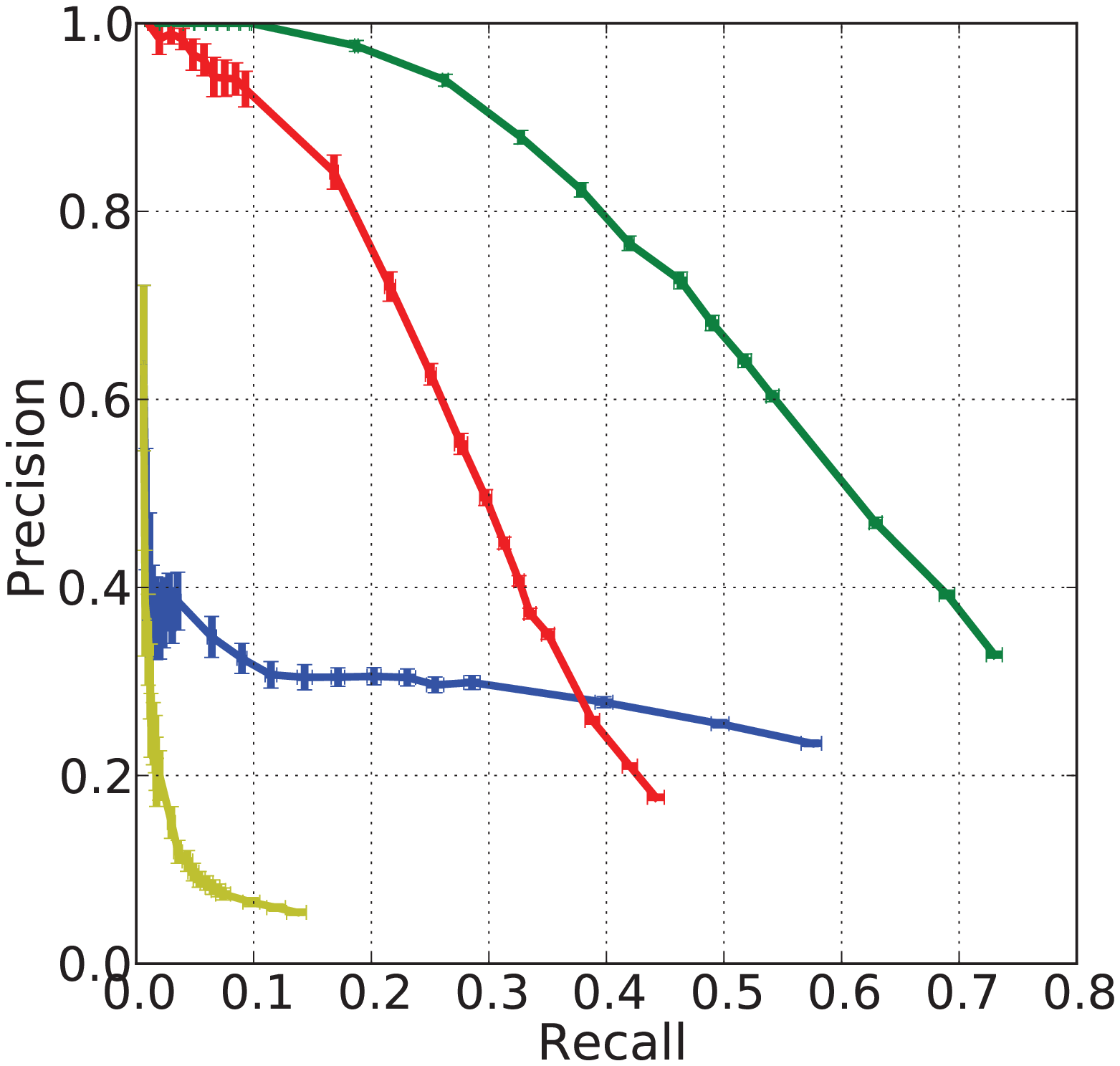}
  }
    \subfloat[][{ROC}]{
    \includegraphics[width=0.225\textwidth]{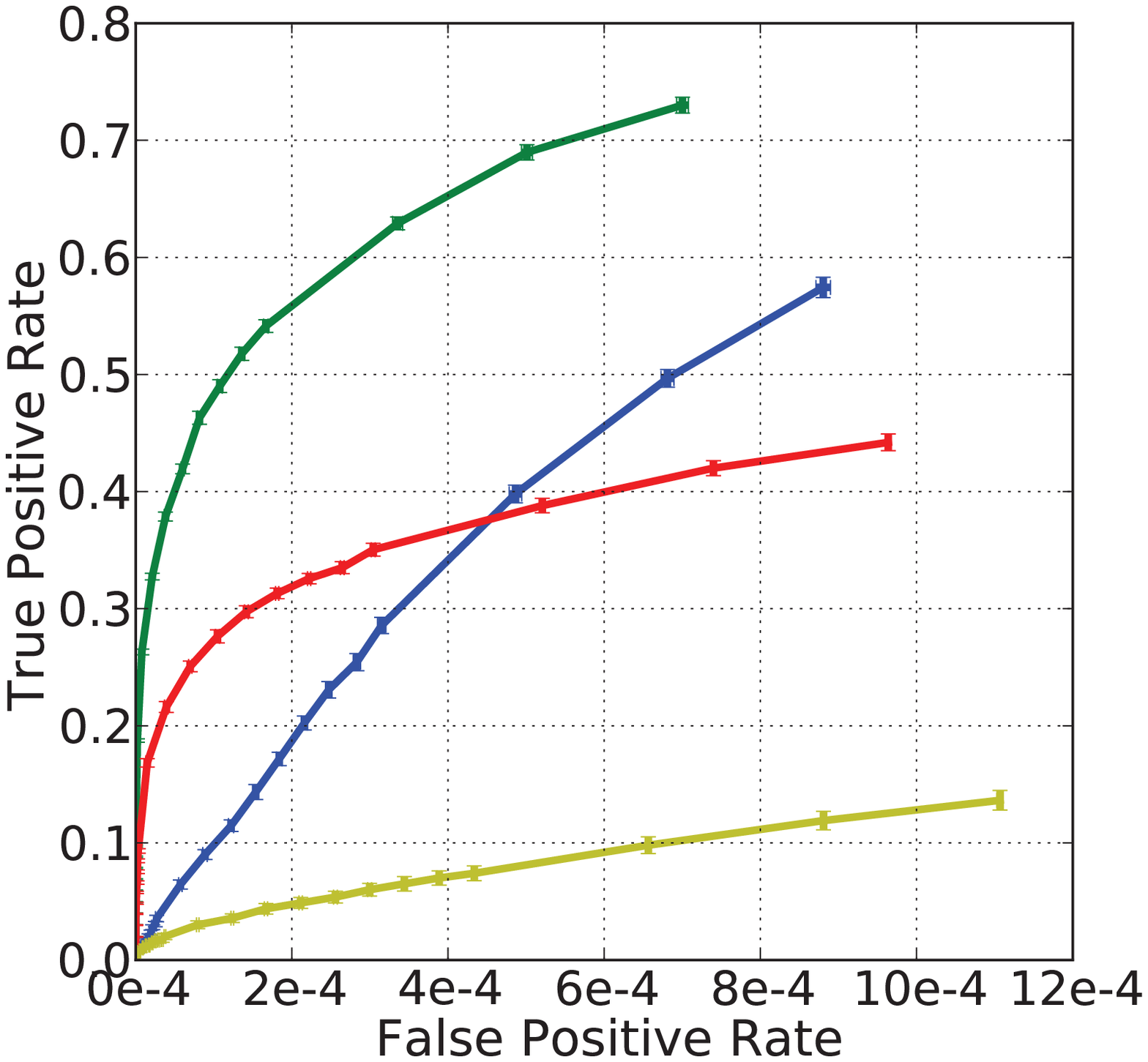}
  }
  \end{center}
        \vspace{-.2cm}
  \caption{
    \textbf{Evaluation of alternative methods on semi-empirical GWAS
      datasets, mimicking population structure as found in
      \emph{Arabidopsis thaliana}}.
    \textbf{(a)}
    Precision-Recall Curve for recovering simulated causal SNPs using alternative methods.
    Shown is precision (TP/(TP+FP)) as a function of the recall (TP/(TP+FN)).
    \textbf{(b)}
    Alternative evaluation of each method on the identical dataset using Receiver
    operating characteristics (ROC).
    Shown is the True Positive Rate (TPR) as a function of the False Positive Rate (FPR).
  }
  \vspace{-.8cm}
  \end{figure}

We assessed the ability of LMM-Lasso to recover true genotype
to phenotype associations in a semi-empirical simulated dataset.
To ensure realistic characteristics of population structure, we
simulated confounding such that it borrows key characteristics from
\textit{Arabidopsis thaliana}, a strongly structured population.

\EDIT{
To compare our method to existing techniques, we considered the
standard Lasso, which models all SNPs jointly but without correcting 
for population structure, as well as univariate Linear Mixed Models,
which effectively control for confounding, but consider each
SNP in isolation.
As a baseline, we also considered a standard univariate Linear Model (LM),
which neither accounts for confounding nor considers joint effects due
to complex genetic architectures.
Both, the standard Lasso and LMM-Lasso were fit in identical ways (See
Section~\ref{sec:model-selection}).
For the linear mixed model and the LMM-Lasso, we used the RRM as
covariance matrix and fit $\delta$ on the null model. 
For univariate models, the ranking of individual SNPs was done
according to their p-values, for multivariate models we considered
the order of inclusion into the model.
A fair comparison between the univariate and multivariate methods is difficult as the univariate methods select blocks of linked markers, whereas the multivariate methods select only one representative marker per block (see Supplementary text S1, Section 1).
}

\paragraph{LMM-Lasso ranks causal SNPs higher than alternative
  methods} \quad \\

 \begin{figure*}[!tb]
   \centering
   \includegraphics[width=0.45\textwidth]{fig2_legend} \\
        \vspace{-0.5cm}
  \subfloat[][Population structure strength]{
    \includegraphics[width=0.3\textwidth]{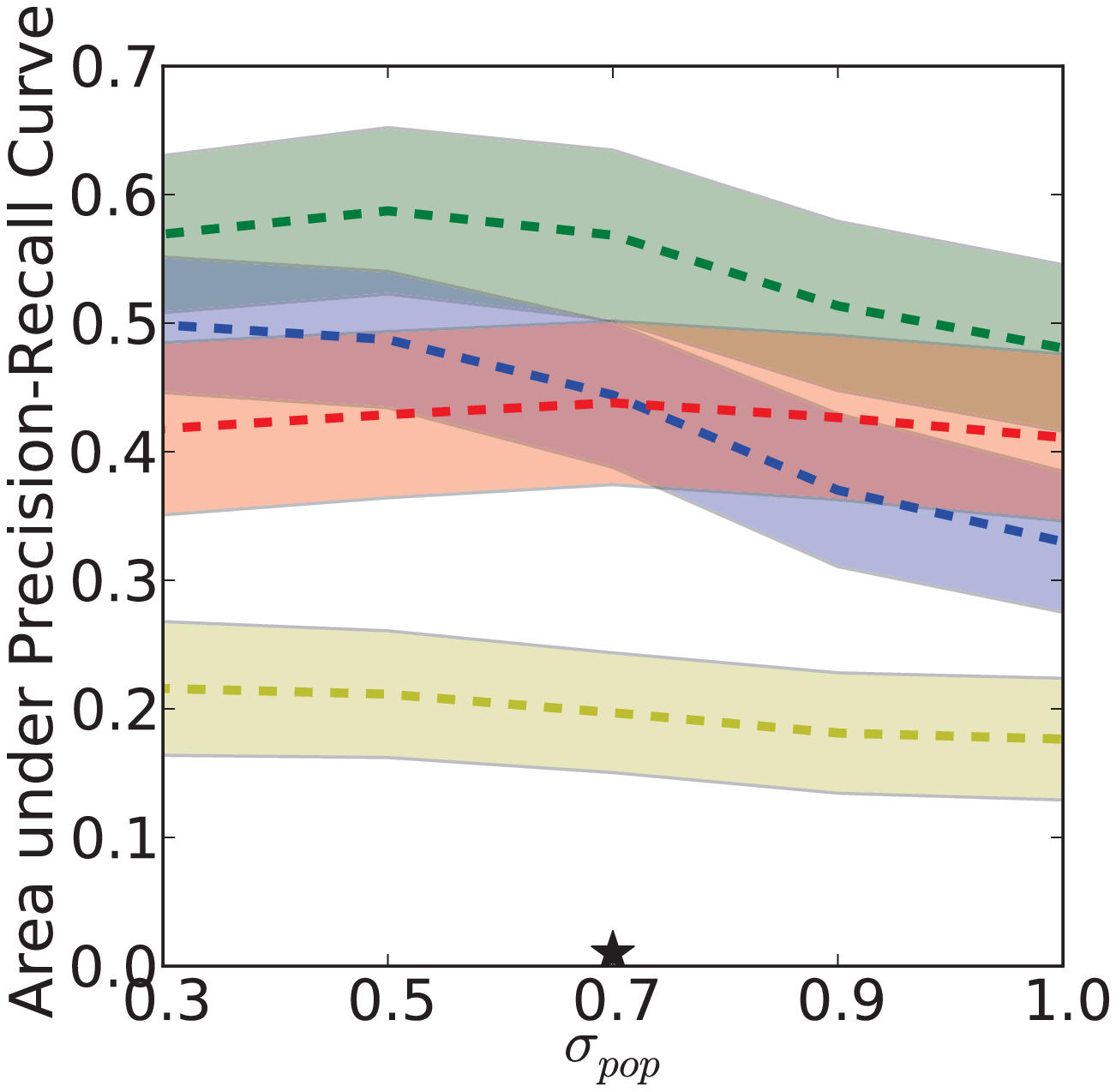}
  }
  \subfloat[][Trait complexity: Varying Number of Causal SNPs]{
      \includegraphics[width=0.3\textwidth]{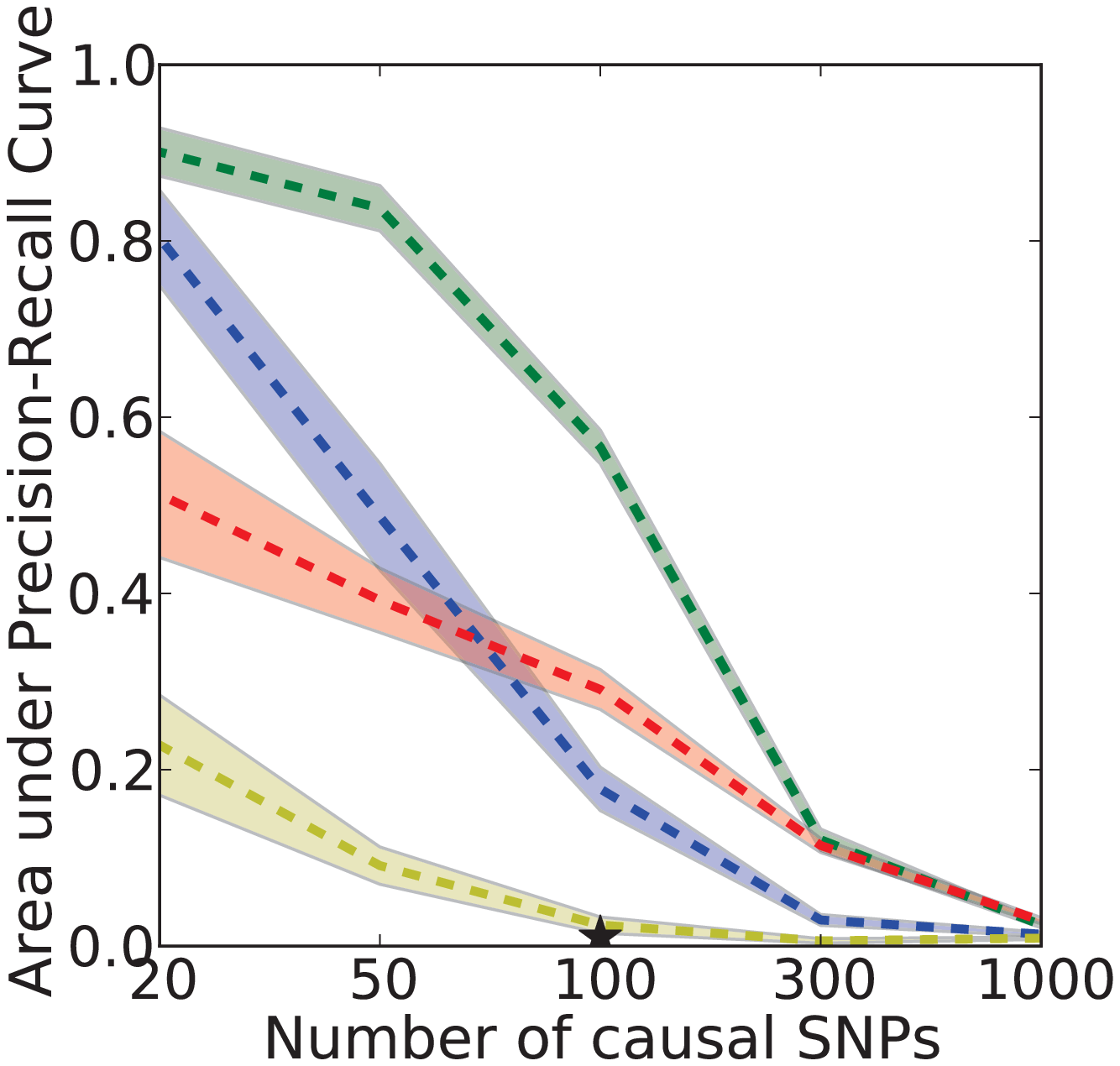}
  }
    \subfloat[][Trait complexity: Varying signal strength]{
    \includegraphics[width=0.3\textwidth]{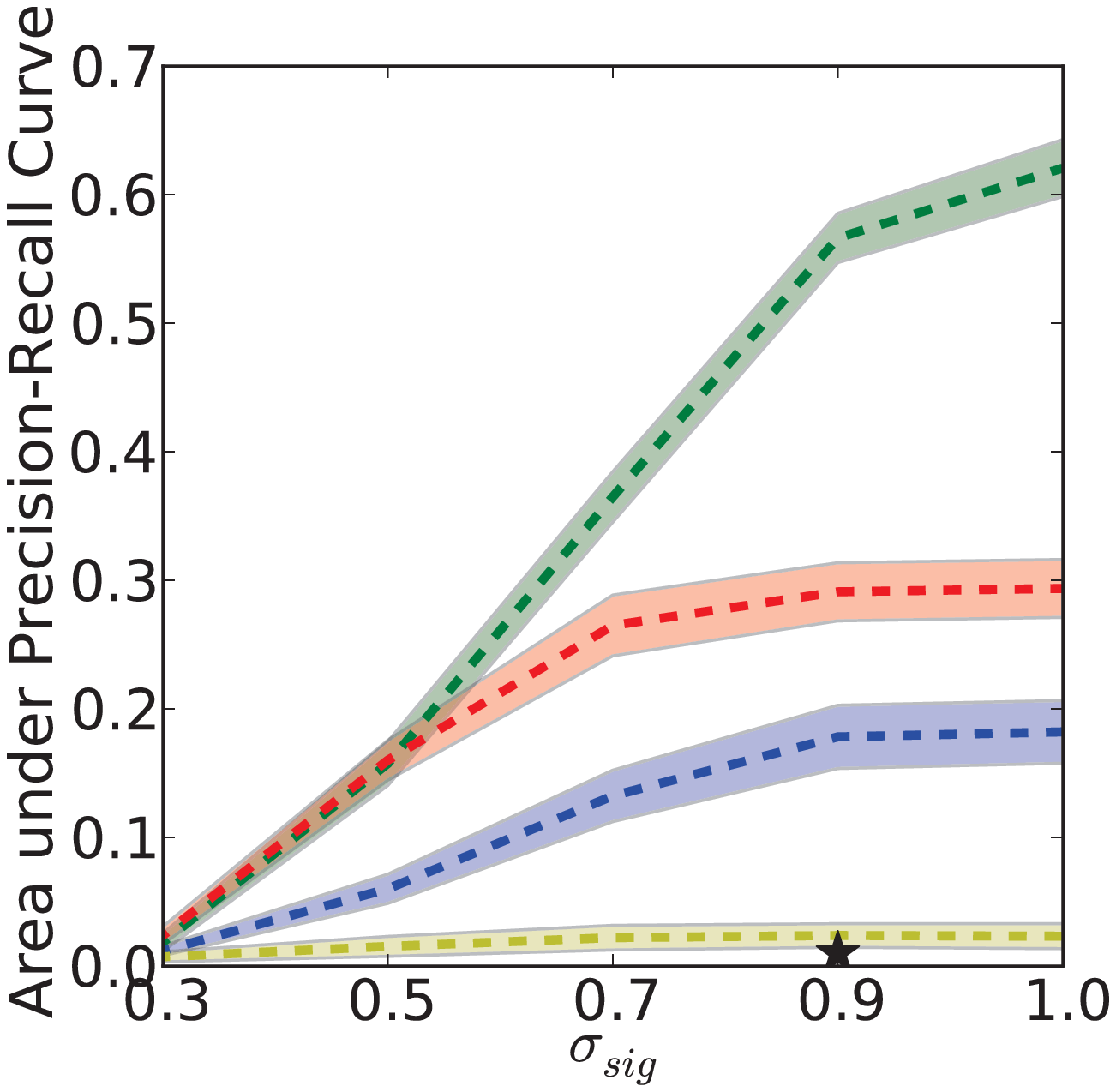}
  } 
          \vspace{-.2cm}
  \caption{
    \textbf{Evaluation of alternative methods on the semi-empirical
      GWAS dataset for different simulation settings.} 
    Area under precision recall curve for finding the true simulated associations.
    Alternative simulation parameters have been varied in a chosen range.
    \textbf{(a)} Evaluation for different relative strength of
    population structure $\sigma_{\text{pop}}$.
    \textbf{(b)} Evaluation for true simulated genetic models with
    increasing complexity (more causal SNPs).
    \textbf{(c)} Evaluation for variable signal to noise ratio
    $\sigma_{\text{sig}}$. 
  }
    \vspace{-.8cm}
\end{figure*}

First, we compared the alternative methods in terms of their accuracy
in recovering SNPs with a true simulated association (Figure 1a).
Methods that account for population structure (LMM-Lasso, LMM) are
more accurate than their counter parts, with LMM-Lasso performing
best. 
While the linear mixed model performs well at recovering strong
associations, the independent statistical testing falls short in detecting
weaker associations which are likely masked by stronger effects (Figure S2a).
Comparing methods that account for population structure and naive
methods, we observe that accounting for this confounding effect
avoids the selection of SNPs that merely reflect relatedness without a
causal effect (Figure S2b).
An alternative evaluation, which considers the receiver operating
characteristic curve, given in Figure 1b, yields identical conclusions.

Next, we explored the impact of variable simulation settings.
\EDIT{
As common in the literature, we used the area under the
precision-recall curve as a summary performance measure to compare
different algorithms.  
Precision and recall both depend on the decision threshold, above which a
marker is predicted to be activated. 
By varying this threshold, one obtains a precision-recall curve.
}
Figure 2a shows the area under the precision recall curve as a
function of an increasing ratio of population structure and
independent environmental noise.
When the confounding population structure is weak, both the Lasso and the
LMM-Lasso perform similar.
As expected, the benefits of population structure correction in
LMM-Lasso are most pronounced in the regime of strong confounding.

We also examined the ability of each method to recover genetic
effects for increasing complexities of the genetic model, varying the
number of true causal SNPs while keeping the overall genetic
heritability fixed (Figure 2b).
\EDIT{
LMM-Lasso performs better than alternative methods for the whole range
of considered settings with the difference in accuracy being the
largest for genetic architectures of medium complexity.
}
In a nutshell, these results show that, in the regime of a larger number of
true weak associations, it is advantageous to include a genetic
covariance $\bfK$ that accounts for some of the weak effects~\cite{visscher-2010-1}.

The identical effect is observed when varying the ratio between true
genetic signal versus confounding and noise (Figure 2c).
\EDIT{
Again, the performance of the LMM-Lasso is superior to all other
methods and the strengths are particularly visible for medium signal to
noise ratios.
}

\subsection{LMM-Lasso explains the genetic architecture of complex
traits in model systems}
Having shown the accuracy of LMM-Lasso in recovering causal SNPs in
simulations, we now demonstrate that the LMM-Lasso better models
the genotype-to-phenotype map in
\textit{Arabidopsis thaliana} and mouse~\cite{valdar2006genome}.
Here, we focus on 20 flowering time phenotypes for
\textit{Arabidopsis thaliana}, which are well characterized,
 and 273 mouse phenotypes which are relevant to human health.

\paragraph{LMM-Lasso more accurately predicts phenotype from genotype
  and uncovers sparser genetic models}
  \begin{figure}[!tb]
     \centering
  \subfloat[][Arabidopsis test variance]
  {
    \includegraphics[width=0.225\textwidth]{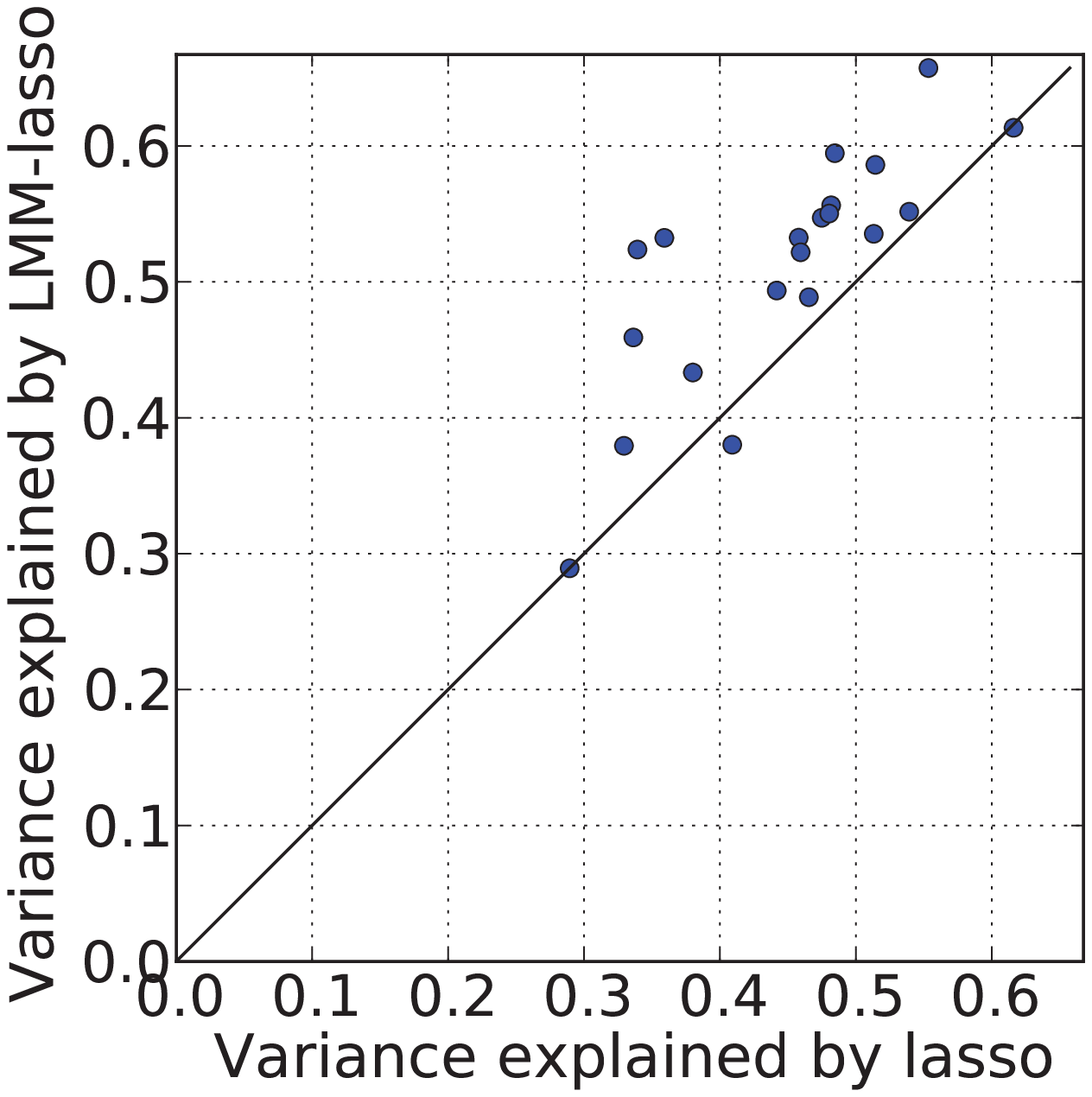}
  }
  \subfloat[][Mouse test variance]
  {
    \includegraphics[width=0.225\textwidth]{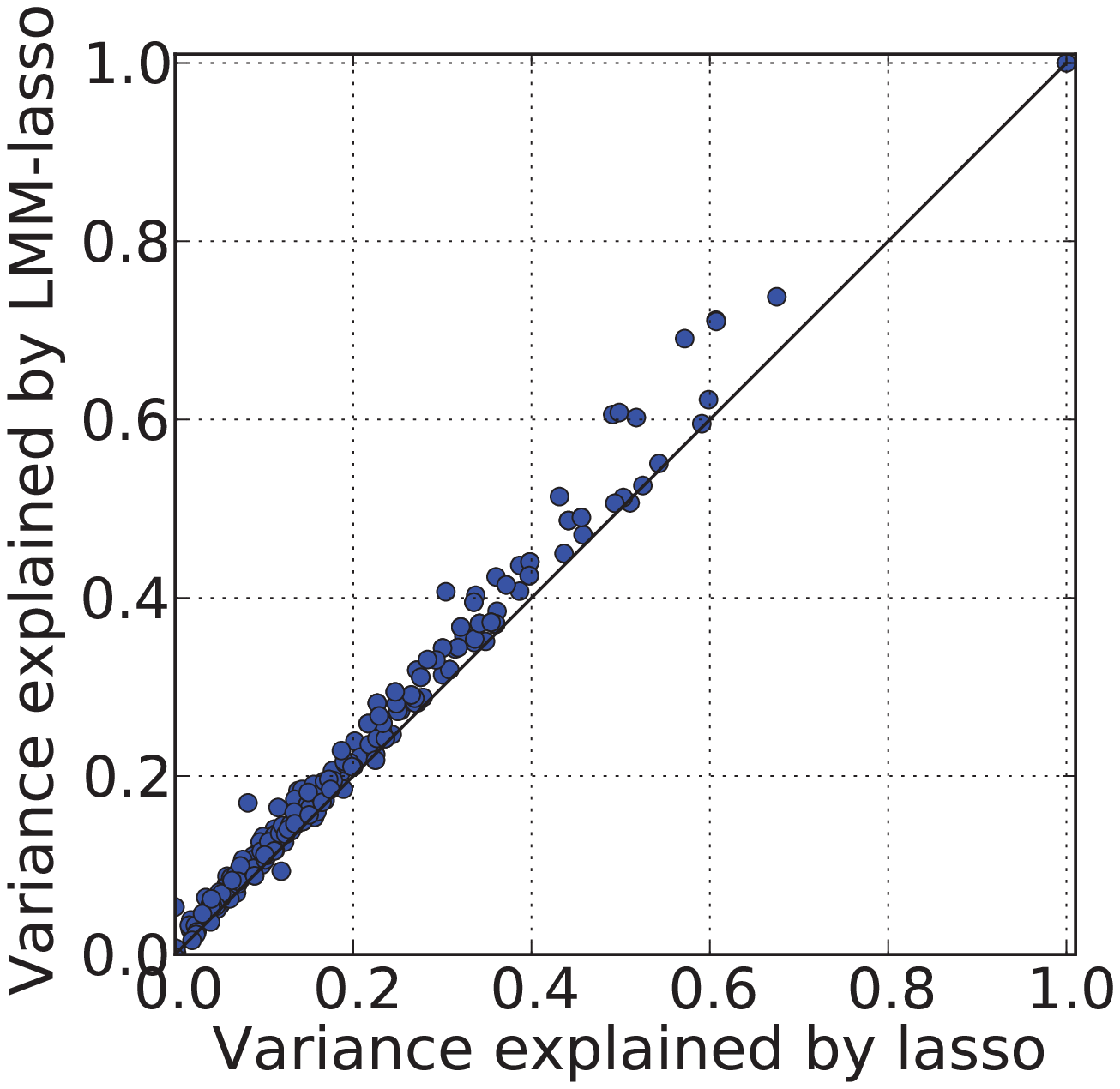}
  }
  \\
  \subfloat[][Arabidopsis number of SNPs]
  {
    \includegraphics[width=0.225\textwidth]{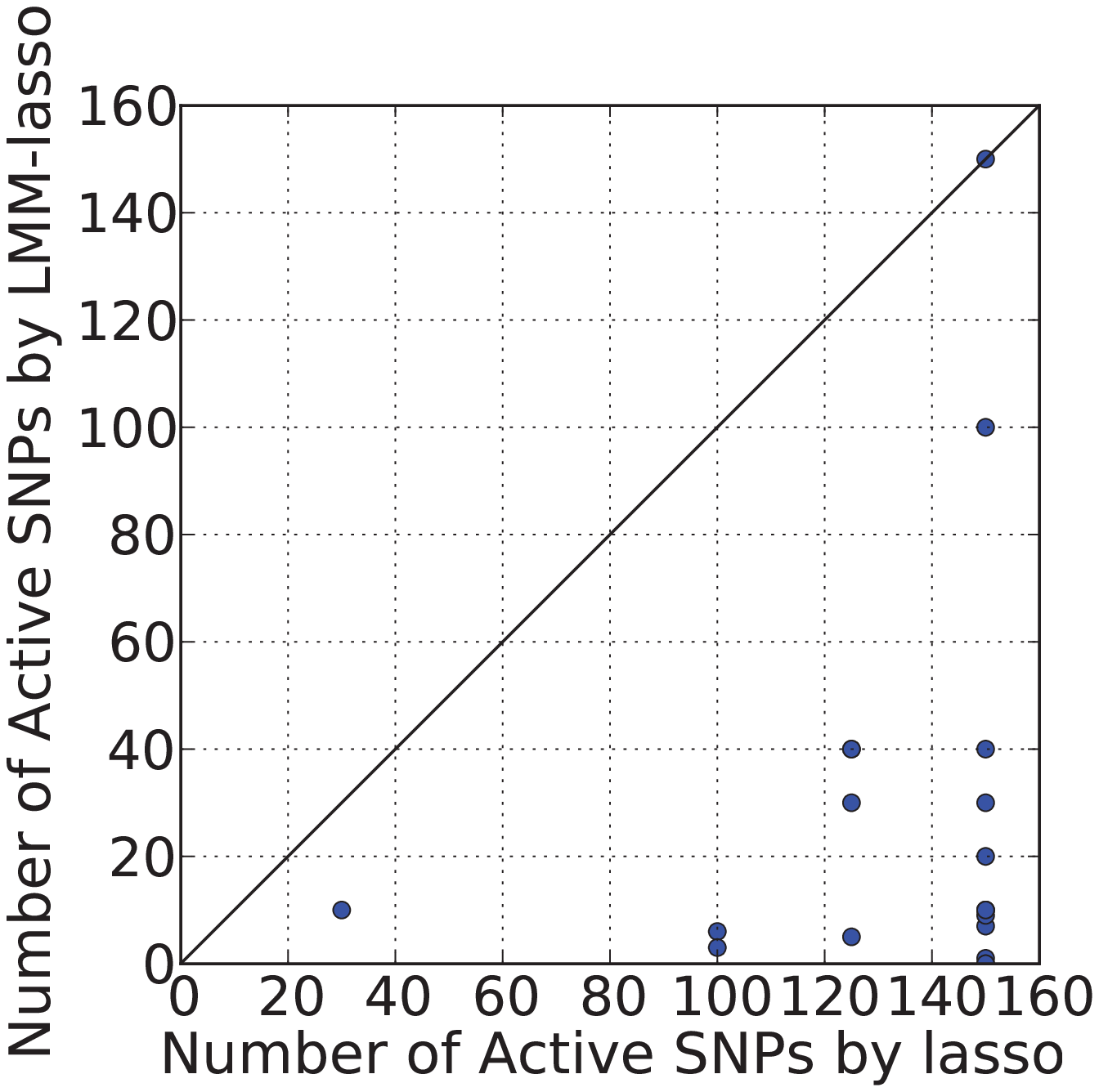}
  }
  \subfloat[][Mouse number of SNPs]
  {
    \includegraphics[width=0.225\textwidth]{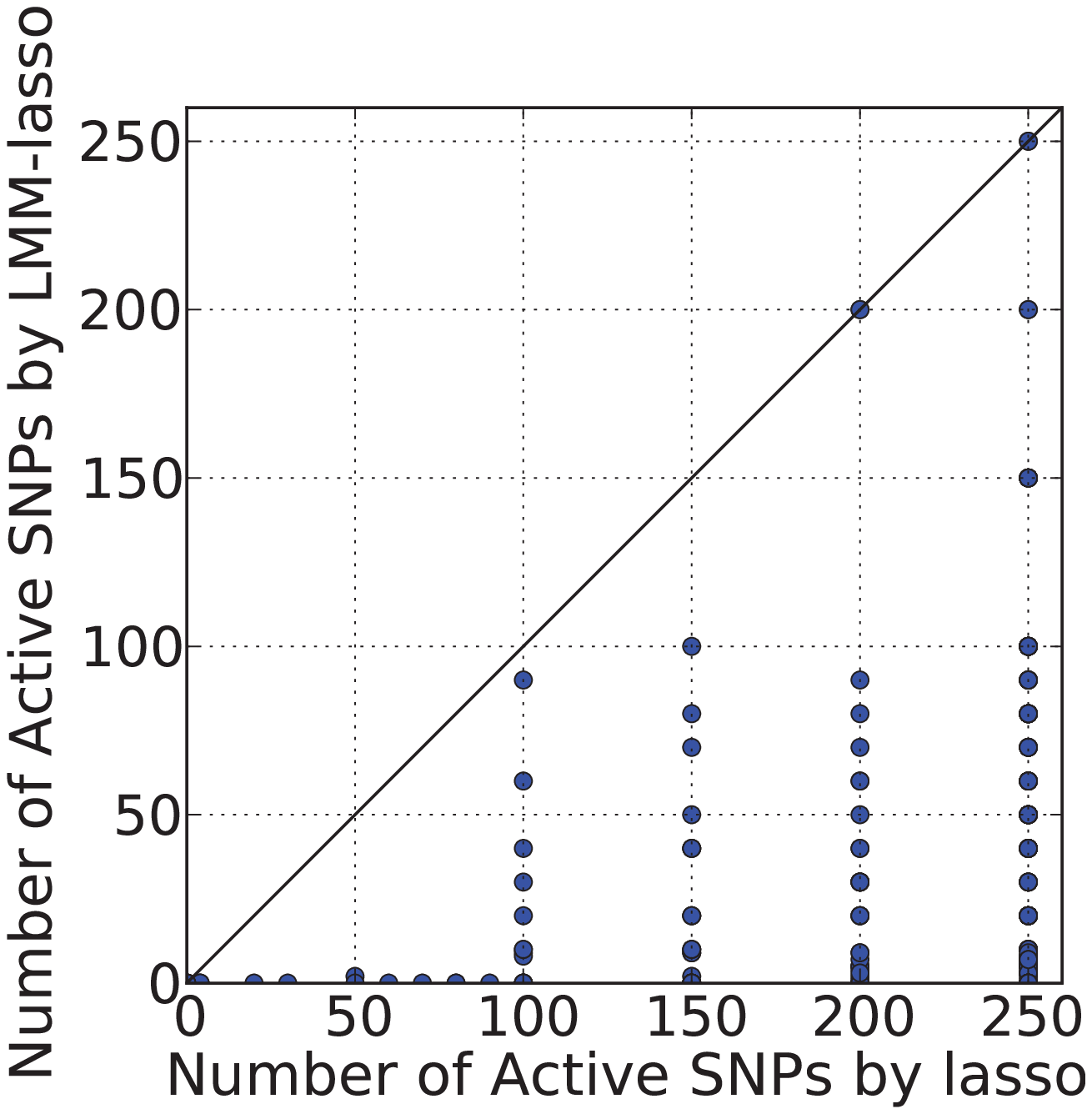}
  }
          \vspace{-.2cm}
  \caption{
    \textbf{
      Predictive power and sparsity of the fitted genetic models for Lasso
      and LMM-Lasso applied to quantitative traits in model systems.
    }
   Considered were flowering phenotypes in \textit{Arabidopsis
      thaliana} and bio-chemical and physiological phenotypes with
    relevance for human health profiled in mouse.
    Comparative evaluations include the fraction of the phenotypic variance
    predicted and the complexity of the fitted genetic model (number
    of active SNPs).
    \textbf{(a)} Explained variance in
    \emph{Arabidopsis}.
    \textbf{(b)} Explained variance in
    mouse.
    \textbf{(c)} Complexity of fitted models in \emph{Arabidopsis}.
    \textbf{(d)} Complexity of fitted models in mouse.
  }
    \vspace{-.8cm}
\end{figure}

First, we considered phenotype prediction to investigate the
capability of alternative methods to explain the joint effect of
groups of SNPs on phenotypes.
To measure for the predictive power, we assessed which
fraction of the total phenotypic variation can be explained by genotype
using different methods~\cite{henner-dros-2012-1}.
Explained variance is defined as the fraction of the total variance of
the phenotype that can be explained by
the model and in our experiments equals one minus the mean squared error as we
preprocesed the data to have zero-mean and unit-variance.
We avoided prediction on the training data, as for all methods
this leads to anti-conservative estimates of variance explained due to
overfitting (see Figure S 4 for a comparison).

Figure 3a and 3b show the explained variance of the two methods on the
independent test data set for each phenotype in the two datasets.
For both model organisms, LMM-Lasso explained at least as much
variation as the Lasso.
We omitted the univariate methods, as their performance is generally
lower due to the simplistic assumption of a single causal SNP (See
Figure S4 for comparative 
predictions in \textit{Arabidopsis thaliana}).
In a fraction of $85.00\%$ of the \textit{Arabidopsis thaliana} and $91.58\%$ of
the mouse phenotypes, LMM-Lasso was more accurate in predicting the
phenotype and thus explained a greater fraction of the
phenotype variability from genetic factors than the Lasso.
In contrast, Lasso achieved better performance in only  $15\%$ of the \textit{Arabidopsis thaliana}
and $8.42\%$ of the mouse phenotypes.
Beyond an assessment of the genetic component of phenotypes, LMM-Lasso
dissects the phenotypic variability into the contributions of individual SNPs and of population structure.
Figure 3c and 3d show the number of SNPs selected in the respective genetic
models for prediction.
With the exception of two phenotypes, LMM-Lasso selected substantially
fewer SNPs than the Lasso, suggesting that the Lasso includes
additional SNPs into the model to capture the effect of population
structure through an additional set of individual SNPs.
This observation is in line with the insights derived from the
simulation setting where the majority of excess SNPs selected by Lasso
are indeed driven by population effects (S 2b).
Although the genetic models fit by LMM-Lasso are substantially
sparser, they nevertheless suggest complex genetic control by multiple
loci.
In $90.00\%$ of \textit{Arabidopsis thaliana} and
in $66.06\%$ of the mouse phenotypes, LMM-Lasso selected more than one
SNP,  in $40.00/45.49\%$ of the cases the number of SNPs in the model
was greater than $10$.

\paragraph{LMM-Lasso allows for dissecting individual SNP
  effects from global genetic effects driven by population structure}
  \begin{figure}[!tb]
  \centering
\includegraphics[width=0.5\textwidth]{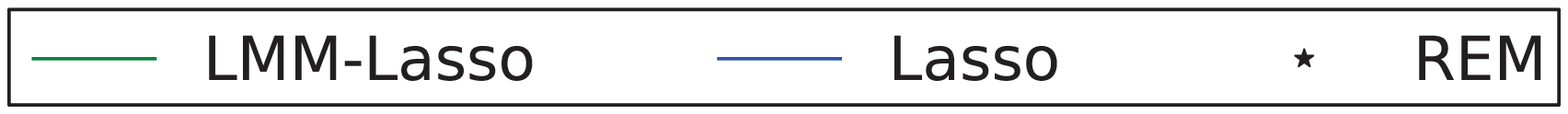} \\
        \vspace{-.2cm}
\includegraphics[width=0.5\textwidth]{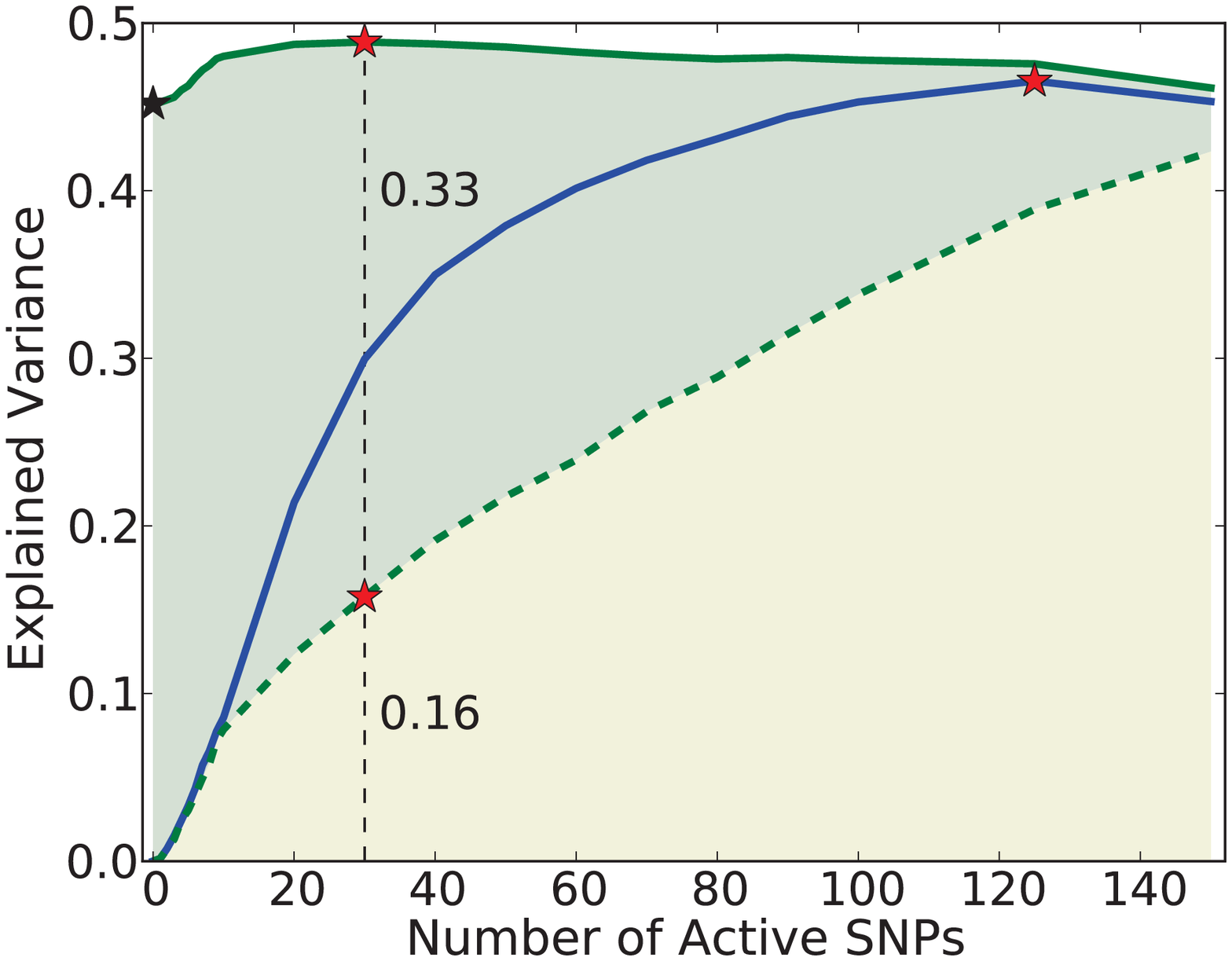}
        \vspace{-.2cm}
  \caption{
  \textbf{
    Variance dissection into individual SNP effects and
    global genetic background driven by population structure.
  }
  Shown is the explained variance on an independent test set as a
  function of the number of active SNPs for the flowering phenotype
  ($10^\circ$) in \textit{Arabidopsis thaliana}. In blue, the
  predictive test set variance of the Lasso as a function of the
  number of SNPs in the model.
  In green, the total predictive variance of LMM-Lasso for different
  sparsity levels.
  The shaded area indicates the fraction of variance LMM-Lasso
  explains by means of population structure (yellow) and population
  structure (green).
  LMM-Lasso without additional SNPs in the model
  corresponds to a genetic random effect model (black star).
}
  \vspace{-.8cm}
\end{figure}

Next, we investigated the ability of LMM-Lasso to differentiate
between individual genetic effects and effects caused by
population structure.
Figure 4 shows the explained variances for the phenotype
flowering time (measured at $10^\circ C$) for \textit{Arabidopsis
thaliana}.
Again, these estimates were obtained using a cross validation approach.
It is known~\cite{zhao_as-2007-1} that flowering is strikingly
associated with population structure, which explains why the LMM-Lasso
already captured a substantial fraction ($45.17\%$) of the phenotypic
variance, when using realized relationships alone (number of active
SNPs=0). 
Due to the small sample size, cross-validation can underestimate the
true explained variance~\cite{hastie_esl-2003-1}.
Nevertheless, cross-validation is fair for comparison and conservative
as it avoids possible overfitting.

For increasing number of SNPs included in the model, the explained
variance of LMM-Lasso gradually shifted from the kernel to the effects
of individual SNPs.
In this example the best performance ($48.87\%$) was reached with
30 SNPs in the model where the relative contribution of the
random effect model was $33.10\%$ and of the individual SNPs is
$15.77\%$. 
In comparison, Lasso explained at most $46.53\%$ of the total variance,
when 125 SNPs were included in the model.

\paragraph{Associations found by LMM-Lasso are enriched for SNPs in
  proximity to known candidate genes}

  \begin{table}[!tb]
\begin{center}
 \begin{tabular}{lrr}
     Phenotype    & LMM-Lasso & Lasso \\
     \hline
LD & \bf{5/54} & 4/69 \\
LDV & \bf{5/63} & 3/69 \\
SD & \bf{3/55} & 2/61 \\
SDV & \bf{5/54} & 2/60 \\
FT10 & 1/48 & \bf{4/67} \\
FT16 & 3/51 & \bf{4/68} \\
FT22 & \bf{2/54} & 1/64 \\
2W & \bf{3/53} & 2/65 \\
8W & 2/51 & \bf{4/59} \\
FLC & \bf{5/52} & 3/53 \\
FRI & 3/43 & 3/46 \\
8WGHFT & \bf{4/59} & 2/66 \\
8WGHLN & 1/48 & \bf{4/58} \\
0WGHFT & \bf{4/58} & 3/63 \\
FTField & \bf{4/61} & 3/69 \\
FTDiameterField & 1/49 & 1/51 \\
FTGH & 1/49 & \bf{2/61} \\
LN10 & \bf{3/50} & 2/67 \\
LN16 & 2/58 & \bf{3/64} \\
LN22 & \bf{4/54} & 2/65 \\
   \end{tabular}
\end{center}

\caption{
{\bf{Associations close to known candidate genes.}}
We report true positives/positives (TP/P) for LMM-Lasso and Lasso for all phenotypes related to flowering time in
 \textit{Arabidopsis thaliana}.
 P are all activated SNPs and TP are all activated SNPs that are close to
 candidate genes.
}
\label{tab:label}
  \vspace{-.8cm}
 \end{table}

Finally, we considered the associations retrieved by alternative
methods in terms of their enrichment near candidate genes with known
implications for flowering in \textit{Arabidopsis thaliana}.
It can be advantageous to remove the SNP of
interest from the population structure covariance (see also discussion
in~\cite{fastlmm}).
Thus, we applied LMM-Lasso on a per-chromosome basis estimating the
effect of population structure from all remaining chromosomes.
To obtain a comparable cutoff of significance, we employed stability
selection for both the LMM-Lasso and Lasso (See
Section~\ref{sec:model-selection}).

Table 1 shows that the LMM-Lasso found a greater number of SNPs
linked to candidate genes for twelve phenotypes, whereas Lasso retrieved a
greater number for only six phenotypes.
In the remaining two phenotypes, both methods performed identically
(For a complete list of candidate genes found by LMM-Lasso, See Table S1).
It is difficult to compare the multivariate approaches with univariate
techniques in a quantitative manner since the univariate models tend
to retrieve complete LD-Blocks. 
Thus, we revert to reporting the p-values of the univariate methods for the
SNPs detected by the LMM-Lasso. 

We also considered to what extent the findings yield evidence for
genetic heterogeneity in proximity to candidate genes (as in the
simulated setting in Figure 3).
Overall, $14.75\%$ of the SNPs linked to candidate genes and selected by the
LMM-Lasso appear as adjacent pairs (Table S2), i.e. having a distance less than 10kb to each other,
while $5.56\%$ of the SNPs selected by the Lasso do.
From all activated SNPs, $8.18\%$ selected by LMM-Lasso and $18.96\%$
selected by the Lasso have at least a second active SNP in close
proximity.
\EDIT{
A simulated example, illustrating how the LMM-Lasso can detect genetic
heterogeneity is shown in Supplementary text S1, Section 3.
}
\end{methods}

\section{Discussion}
Here, we have presented a Lasso multi-marker mixed model (LMM-Lasso) for
detecting genetic associations in the presence of confounding
influences such as population structure. 
The approach combines the attractive properties of mixed models that
allow for elegant correction for confounding effects and those of
multi-marker models that consider the joint effects of sets of genetic
markers rather than one single locus.
As a result, LMM-Lasso is able to better recover true genetic effects, even in
challenging settings with complex genetic architectures, weak effects
of individual markers or presence of strong confounding effects.

LMM-Lasso is relevant for genome-wide association studies of complex
phenotypes, particularly the large number of phenotypes whose genetic
basis is conjectured to be multifactorial~\cite{flint-2009-1}.
Here, we have demonstrated such practical use through retrospective
analysis of  \emph{Arabidopsis thaliana} and data from inbred mouse
lines.  
First, we found that the combination of random effect modeling and  
multivariate linear models as done in LMM-Lasso improves the
prediction of phenotype from genotype,
suggesting that the underlying model that accounts for both,
population structure effects and multi-locus effects, is a better fit
to real genetic architectures.
It is widely accepted that the missing heritability in single-locus
genome wide association mapping can often be explained by a large
number of loci that have a joint effect on the
phenotype~\cite{visscher-2010-1} while leading only to weak signals of
association if considered independently.
In addition to recovering greater fractions of the heritable component
of quantitative traits, LMM-Lasso allows for differentiating between
variation that is broad-scale genetic and hence likely caused by
population structure and individual genetic effects.
In \emph{Arabidopsis} and mouse, this approach revealed substantially
sparser genetic models than naive Lasso approaches.
Second, LMM-Lasso retrieves genetic associations that are enriched for known
candidate genes.
In line with the findings in~\cite{yang2012conditional}, we retrieved
an increased rate of physically adjacent SNPs selected in proximity to
candidate genes.

\EDIT{
Neither the concept of accounting for population structure nor
multivariate modeling of the genetic data are novel \emph{per se}.
An approach for distincting populations based on multi-task learning is
presented in ~\cite{xing-2010-1}.
There is a vast amount of literature using a $\ell_1$-regularized
approach for genome-wide associations
studies~\cite{wu_gwaslars-2009-1,xing-2012, kim2009}. 
In \cite{foster-2007-1}, as sparse random effect model is proposed, 
where the markers are modeled as random Lasso effects.
In ~\cite{balding-2008-1,li2011bayesian}, the authors suggest to add
principal components to the model to correct for population
structure. 
While these approaches can be effective in some settings, principal
components cannot account for family structure or cryptic
relatedness~\cite{price2010}.  
Importantly, none of these approaches considers including random effects
to control for confounding. 
A notable exception is the general L1 mixed model framework by
Schelldorfer et. al.~\cite{schelldorfer-2011-1}, who consider a
random effect component but do not provide a scalable algorithm that
is applicable to genome-wide settings.

The proposed model is also closely related to existing mixed models,
however these are predominantly considering individual SNPs in
isolation. 
An exception is work in parallel~\cite{bjarni2012} who propose a joint
model of multiple large effect loci in a mixed model using a step-wise
regression approach. 
An important difference to our work is the sequential selection of
SNPs, which implies an effect due to ordering whereas LMM-Lasso selects
all SNPs jointly. 
}

As sample sizes increase, the power of detecting multifactorial effects
will quickly rise.
Moreover, larger datasets improve the feasibility to estimate accurate
p-values of individual markers by using stability
selection~\cite{meinshausen-2009-1}, which involves randomized
splitting of the dataset.
\EDIT{
However, it is unclear how strongly the sample size splitting affects
the power of Lasso-based methods.
Our results suggest that $\ell_1$-regularized methods can indeed be an
attractive tool for fitting multifactorial effects in genetic 
settings, however assessing the statistical significance without
loosing power remains a challenge for future research for Lasso
methods in general.
} 

LMM-Lasso addresses the problem that multi-marker mapping is
inherently linked to the challenge of some markers being picked up by
the model due to their correlation with a confounding variable, such
as population structure. 
In a pure Lasso regression model, it is unclear which markers merely
reflect these hidden confounders.
LMM-Lasso on the other hand explains confounding explicitly as random
effect, and thus, helps to resolve the ambiguity between individual
genetic effects and phenotype variability due to population
structure.
In summary, we therefore deem the LMM-Lasso a useful addition to the
current toolbox of computational models for unraveling
genotype-phenotype relationships.

\section*{Acknowledgement}
The authors would like to thank Bjarni J. Vilhjalmsson and Yu Huang
for providing the list of genes that are involved in flowering of
\emph{A. thaliana}, and Nicolo Fusi for preprocessing of the mouse data.

\paragraph{Funding\textcolon}
B.R., C.L. and K.B. were funded by the Max Planck Society.
O.S. was supported by a Marie Curie FP7 fellowship.

\bibliographystyle{abbrv}
\bibliography{bibfile}

\end{document}

%% file: utils.tex
%
%
%
%
%

\newcommand{\EDIT}[1]{{\color{black} #1}}

\newcommand{\mbf}[1]{{\ensuremath{\mathbf{#1}}}}

\newcommand{\data}{\mathcal{D}}
\newcommand{\model}{\mathcal{H}}
\newcommand{\GPM}{\mathcal{H}_{\text{GP}}}
\newcommand{\datatest}{\mathcal{D}_{\text{test}}}
\newcommand{\pl}{\ensuremath{p_{\textnormal{L}}}}
\newcommand{\TK}{\ensuremath{\bTheta_{\textnormal{K}}}}
\newcommand{\hTL}{\ensuremath{\hat{\btheta}_{\textnormal{L}}}}
\newcommand{\hTK}{\ensuremath{\hat{\bTheta}_{\textnormal{K}}}}
\newcommand{\TL}{\ensuremath{\btheta_{\textnormal{L}}}}
\newcommand{\TT}{\ensuremath{\btheta}}
\newcommand{\x}{\ensuremath{\bfx}}
\newcommand{\X}{\ensuremath{\bfX}}

\newcommand{\argmax}{\operatornamewithlimits{argmax}}

\newcommand{\rmd}{\mathrm{d}}

\newcommand{\R}{{\sf R\hspace*{-0.9ex}\rule{0.15ex}%
    {1.5ex}\hspace*{0.9ex}}}
\newcommand{\N}{{\sf N\hspace*{-1.0ex}\rule{0.15ex}%
    {1.3ex}\hspace*{1.0ex}}}
\newcommand{\Q}{{\sf Q\hspace*{-1.1ex}\rule{0.15ex}%
    {1.5ex}\hspace*{1.1ex}}}
\newcommand{\C}{{\sf C\hspace*{-0.9ex}\rule{0.15ex}%
    {1.3ex}\hspace*{0.9ex}}}

\newcommand{\be}{\begin{equation}}
\newcommand{\ee}{\end{equation}}
\newcommand{\bea}{\begin{eqnarray}}
\newcommand{\eea}{\end{eqnarray}}
\newcommand{\beas}{\begin{eqnarray*}}
\newcommand{\eeas}{\end{eqnarray*}}

\newcommand{\const}{{\rm const.}}
\newcommand{\matlab}{${\rm Matlab}^{\Pisymbol{psy}{226}}$}

\newcommand{\HGP}{\ensuremath{\mathcal{H}_{\textnormal{GP}}}}

\newcommand{\comment}[1]{}

\newcommand{\TODO}[1]{{\color{red}\fbox{TODO} #1}}

\newcommand{\normal}[2]{\mathcal{N}(#1 \given #2)}

\newcommand{\indep}{\bot \hspace{-0.6em} \bot}
\newcommand{\arrow}{\rightarrow}
\newcommand{\given}{\,|\,}
\newcommand{\twolines}{\,||\,}
\newcommand{\narroweq}{\!\!=\!\!}

\newcommand{\pa}[1]{{\rm pa_\mathit{#1}}}
\newcommand{\cp}[2]{{\rm cp_\mathit{#1}^{(\mathit{#2})}}}
\newcommand{\ch}[1]{{\rm ch_\mathit{#1}}}

\newcommand{\neigh}[1]{{\rm ne_\mathit{#1}}}

\newcommand{\KL}{{\rm KL}}

\newcommand{\entropy}{{\mathbb{H}}}

\newcommand{\cip}{\mbox{$\perp\!\!\!\perp$}}
\newcommand{\condindep}[3]{#1~\cip~#2~|~#3}
\newcommand{\nocondindep}[3]{#1~\mbox{$\not\!\perp\!\!\!\perp$}~#2~|~#3}
\newcommand{\dir}[2]{{\rm Dir}(#1|#2)}

\newcommand{\bDelta}{\mbox{\boldmath $\Delta$}}
\newcommand{\bbeta}{\mbox{\boldmath $\beta$}}
\newcommand{\bmu}{\mbox{\boldmath $\mu$}}
\newcommand{\bnu}{\mbox{\boldmath $\nu$}}
\newcommand{\balpha}{\mbox{\boldmath $\alpha$}}
\newcommand{\bepsilon}{\mbox{\boldmath $\epsilon$}}
\newcommand{\bgamma}{\mbox{\boldmath $\gamma$}}
\newcommand{\bsigma}{\mbox{\boldmath $\sigma$}}
\newcommand{\bSigma}{\mbox{\boldmath $\Sigma$}}
\newcommand{\btau}{\mbox{\boldmath $\tau$}}
\newcommand{\blambda}{\mbox{\boldmath $\lambda$}}
\newcommand{\bLambda}{\mbox{\boldmath $\Lambda$}}
\newcommand{\bpi}{\mbox{\boldmath $\pi$}}
\newcommand{\bpsi}{\mbox{\boldmath $\psi$}}
\newcommand{\bchi}{\mbox{\boldmath $\chi$}}
\newcommand{\bxi}{\mbox{\boldmath $\xi$}}
\newcommand{\bPsi}{\mbox{\boldmath $\Psi$}}
\newcommand{\bphi}{\mbox{\boldmath $\phi$}}
\newcommand{\bPhi}{\mbox{\boldmath $\Phi$}}

\newcommand{\btheta}{\mbox{\boldmath $\theta$}}
\newcommand{\bTheta}{\mbox{\boldmath $\Theta$}}
\newcommand{\bOmega}{\mbox{\boldmath $\Omega$}}

\newcommand{\Bmath}[1]{\mbox{\boldmath $#1$}}

\newcommand{\fastfig}[4]{
\begin{center}
\begin{figure}[htb!]
\centerline{\epsfig{figure=#1,width=#2}}
\caption[short]{#3}
\label{#4}
\end{figure}
\end{center}
}

\newcommand{\unit}{{\bf I}}
\newcommand{\boldzero}{{\bf 0}}

\newcommand{\bfa}{{\bf a}}
\newcommand{\bfb}{{\bf b}}
\newcommand{\bfc}{{\bf c}}
\newcommand{\bfd}{{\bf d}}
\newcommand{\bfe}{{\bf e}}
\newcommand{\bff}{{\bf f}}
\newcommand{\bfg}{{\bf g}}
\newcommand{\bfh}{{\bf h}}
\newcommand{\bfi}{{\bf i}}
\newcommand{\bfk}{{\bf k}}
\newcommand{\bfl}{{\bf l}}
\newcommand{\bfm}{{\bf m}}
\newcommand{\bfp}{{\bf p}}
\newcommand{\bfr}{{\bf r}}
\newcommand{\bfs}{{\bf s}}
\newcommand{\bft}{{\bf t}}
\newcommand{\bfu}{{\bf u}}
\newcommand{\bfv}{{\bf v}}
\newcommand{\bfw}{{\bf w}}
\newcommand{\bfx}{{\bf x}}
\newcommand{\bfy}{{\bf y}}
\newcommand{\bfz}{{\bf z}}

\newcommand{\bfA}{{\bf A}}
\newcommand{\bfB}{{\bf B}}
\newcommand{\bfC}{{\bf C}}
\newcommand{\bfD}{{\bf D}}
\newcommand{\bfG}{{\bf G}}
\newcommand{\bfH}{{\bf H}}
\newcommand{\bfI}{{\bf I}}
\newcommand{\bfJ}{{\bf J}}
\newcommand{\bfK}{{\bf K}}
\newcommand{\bfL}{{\bf L}}
\newcommand{\bfM}{{\bf M}}
\newcommand{\bfQ}{{\bf Q}}
\newcommand{\bfR}{{\bf R}}
\newcommand{\bfS}{{\bf S}}
\newcommand{\bfT}{{\bf T}}
\newcommand{\bfU}{{\bf U}}
\newcommand{\bfV}{{\bf V}}
\newcommand{\bfW}{{\bf W}}
\newcommand{\bfX}{{\bf X}}
\newcommand{\bfY}{{\bf Y}}
\newcommand{\bfZ}{{\bf Z}}
\newcommand{\llangle}{{\langle \hspace{-0.7mm} \langle}}
\newcommand{\rrangle}{{\rangle \hspace{-0.7mm} \rangle}}
\newcommand{\define}{\stackrel{\mathrm{def}}{=}}

\newcommand{\la}{\langle}
\newcommand{\ra}{\rangle}
\newcommand{\La}{\left\langle}
\newcommand{\Ra}{\right\rangle}
\newcommand{\EXP}[1]{\left\langle #1 \right\rangle}
\newcommand{\vectwo}[2]{\left[\begin{array}{c} #1 \\ #2 \end{array}\right]}
\newcommand{\vecn}[1]{\left[\begin{array}{c} #1 \end{array}\right]}
\newcommand{\half}{{\scriptstyle \frac{1}{2}}}
\newcommand{\col}{\mathrm{vec}}

\newcommand{\trans}[1]{{#1}^{\ensuremath{\mathsf{T}}}}
\newcommand{\T}{{\rm T}}
\newcommand{\diag}{{\rm diag}}
\newcommand{\Tr}{\mbox{Tr}}
\newcommand{\diff}[1]{{\,d#1}}
\newcommand{\vgraph}[1]{
  \newpage
  \begin{center}
  {\large \bf #1}
  \end{center}
  \vspace{2mm}
}
\newcommand{\high}[1]{\textcolor{blue}{\emph{#1}}}
\newcommand{\cut}[1]{}
\newcommand{\citeasnoun}[1]{\citeN{#1}}
\newcommand{\citemulti}[2]{(#1, \citeyearNP{#2})}
\newcommand{\citemultiN}[2]{#1 (\citeyearNP{#2})}
\newcommand{\Sum}{{\displaystyle \sum}}
